\documentclass[11pt]{article}
\usepackage{moriond,epsfig}

\def\B{\mathrm{B}}

\def\epem{\mathrm{e^{+}e^{-}}}

\def\etal{\it et~al.}

\def\pp{\mathrm{p\bar{p}}}

\def\qbar{\mathrm{\bar{q}}}
\def\qqbar{\mathrm{q\bar{q}}}

\def\thfs {\mathrm{\theta_{12}}}
\def\thst {\mathrm{\theta_{23}}}
\def\thtf {\mathrm{\theta_{31}}}

\def\Pni{\prod_{i=0}^n{\rm P}_i}

\def\be{\begin{equation}}
\def\ee{\end{equation}}
\def\bea{\begin{eqnarray}}
\def\eea{\end{eqnarray}}

\begin{document}
\vspace*{1cm}
\title{STUDY OF RAPIDITY GAP EVENTS IN HADRONIC Z DECAYS
 WITH THE L3 DETECTOR \footnote{Talk presented at the XXXVII Rencontre de Moriond (QCD),
 Les Arcs, March 2002}}

\author{J.H.FIELD}

\address{D\'{e}partement de Physique Nucl\'{e}aire et Corpusculaire
 Universit\'{e} de Gen\`{e}ve . 24, quai Ernest-Ansermet
 CH-1211 Gen\`{e}ve 4.}

\maketitle\abstracts{
 A search is performed in symmetric 3-jet hadronic Z decay events for evidence 
 of colour singlet exchange and for colour reconnection
 effects predicted by the Rathsman model. Asymmetries in particle flow
 and the angular separation of particles are found to be sensitive to such effects.
 95$\%$ upper limits of 7-9$\%$ are found for the fraction of colour singlet exchange,
 and of 0.93$\%$ for the colour reconnection parameter $R_0$ (default value 0.1) of
 the Rathsman model.}

\section{Introduction}

\begin{figure}[htbp]
\begin{center}
\includegraphics[width=.49\textwidth]{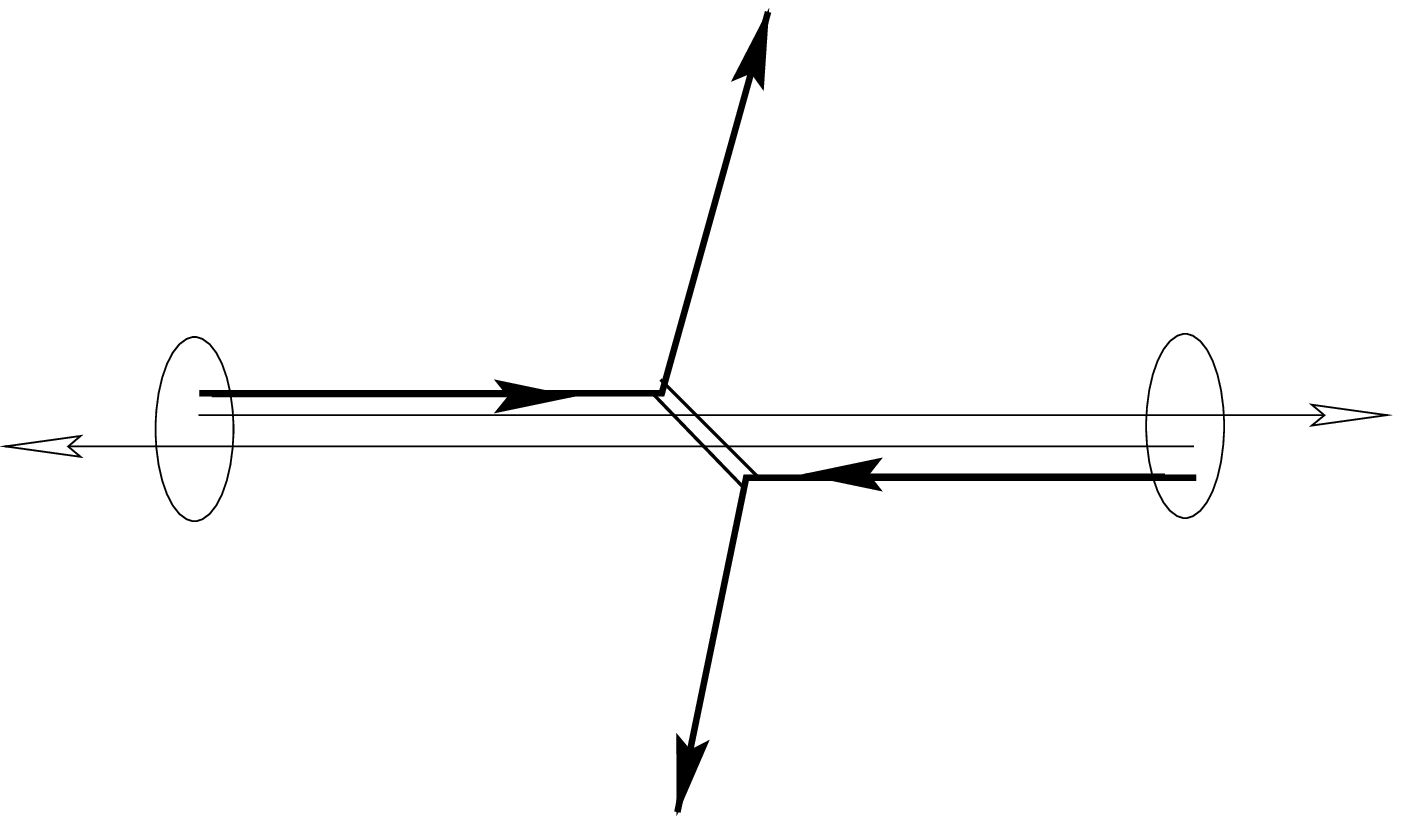}
\hfill
\includegraphics[width=.49\textwidth]{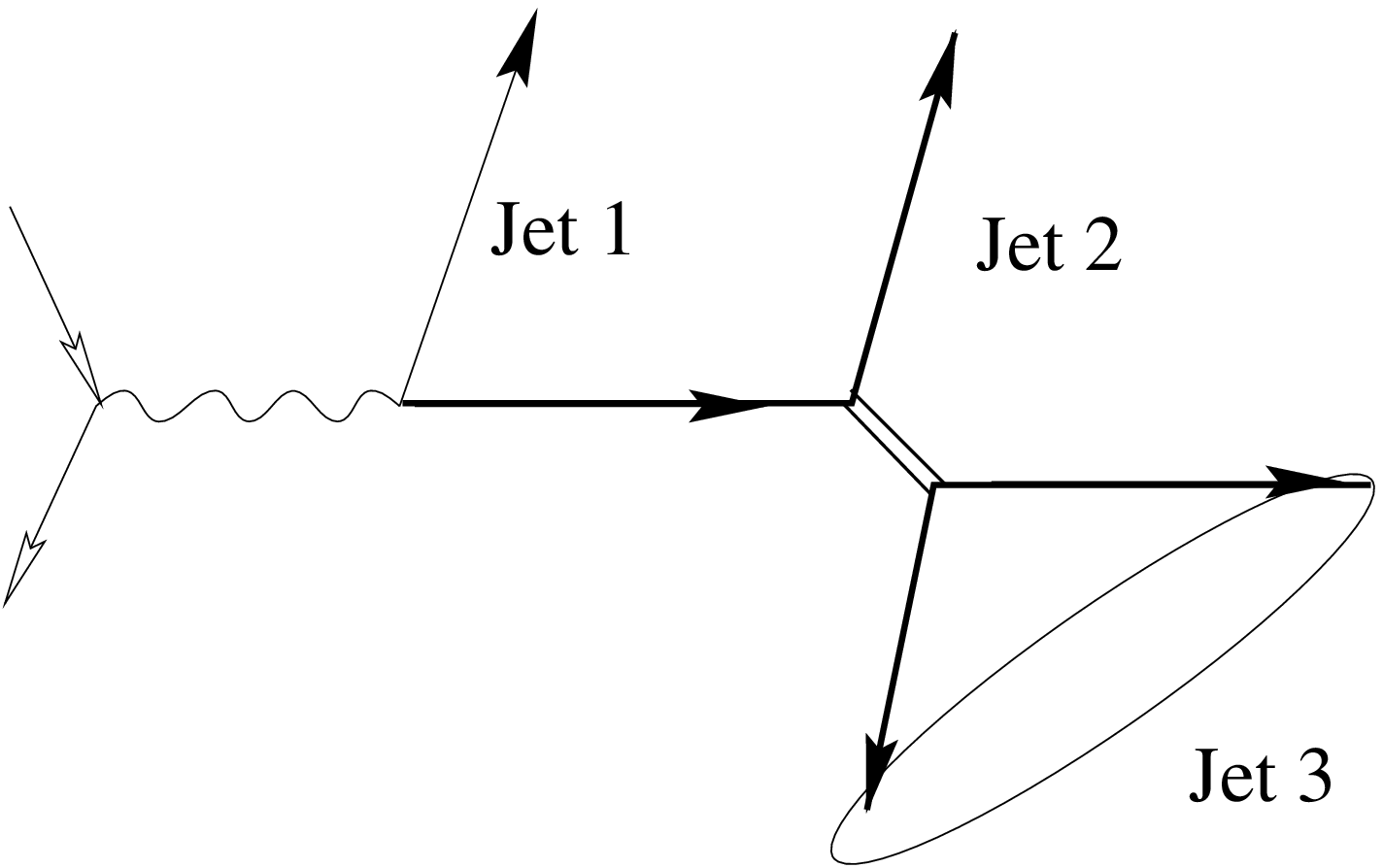}
\caption[]{{\textsc{Cse}} in $\pp$ (left) and in $\epem$ (right) reactions 
shown by double lines.}
\label{fig:cross}
\end{center}
\end{figure}

\begin{figure}[htbp]
\begin{center}
\includegraphics[width=.5\textwidth]{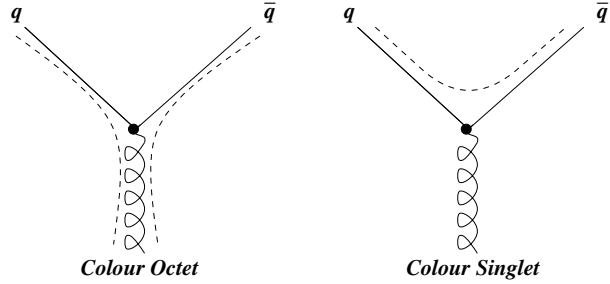}
\caption[]{Colour flow in \textsc{Coe} (left) 
 and \textsc{Cse} (right) shown by dotted lines.}
 \label{fig:cse_coe}
\end{center}
\end{figure}

Events with large rapidity gaps, attributed to 
colour singlet exchange (\textsc{Cse}),
have been observed at {\textsc{Hera}}~\cite{hera} 
and the {\textsc{Tevatron}}~\cite{tevatron}.
By crossing symmetry, we may expect similar gaps in three 
jet hadronic Z decays (figure~\ref{fig:cross}).
Large rapidity gaps have been observed in 1-2$\%$ and $\simeq$10$\%$ of events with two
 high $p_t$ jets at the {\textsc{Tevatron}} and {\textsc{Hera}} respectively.
 The `gap survival probability' due to overlap with particles produced
 in the underlying event has been estimated~\cite{survprob} to be about 20$\%$ at the
 {\textsc{Tevatron}}. An advantage of the Z decay study is the absence of
this supression factor as there is no underlying
  event.  
This study, performed by L3~\cite{l3note}, searches for such gaps by exploiting
differences in colour flow, in 3-jet events, 
between \textsc{Cse} and colour octet exchange (\textsc{Coe}):
in the latter colour flow is present between the qg and $\qbar$g gaps and 
is inhibited by destructive interference in the $\qqbar$ gap,
while in the former case colour flow occurs predominantly in the $\qqbar$ gap  
(figure~\ref{fig:cse_coe}).
For this study, the \textsc{Jetset} Parton Shower program
~\cite{jetset} 
has been used to model \textsc{Coe}.
Two simple models have been used to simulate the expected colour flow in 
\textsc{Cse}:
events of type $\qqbar\gamma$ with a photon effective mass as in the gluon 
jet mass distribution are generated, and the photon is then replaced 
by a boosted quark dijet (model \textsc{CS0}), or 
by a gluon fragmenting independently (model \textsc{CS2}).
The \textsc{Rathsman}~\cite{rath} model, tuned~\cite{tuning} to L3 hadronic Z decay data
\footnote{A preliminary tuning
 has been performed for three different values of the colour recombination
 parameter: $R_0 = 0.037, 0.1, 0.2$. The default value of 0.1 was 
 obtained by fitting the model to H1 data on the diffractive proton 
 structure function~\cite{rath}.}
, is also studied.
\section{Methodology}

\begin{figure}[htbp]
\begin{center}
\includegraphics[width=.35\textwidth]{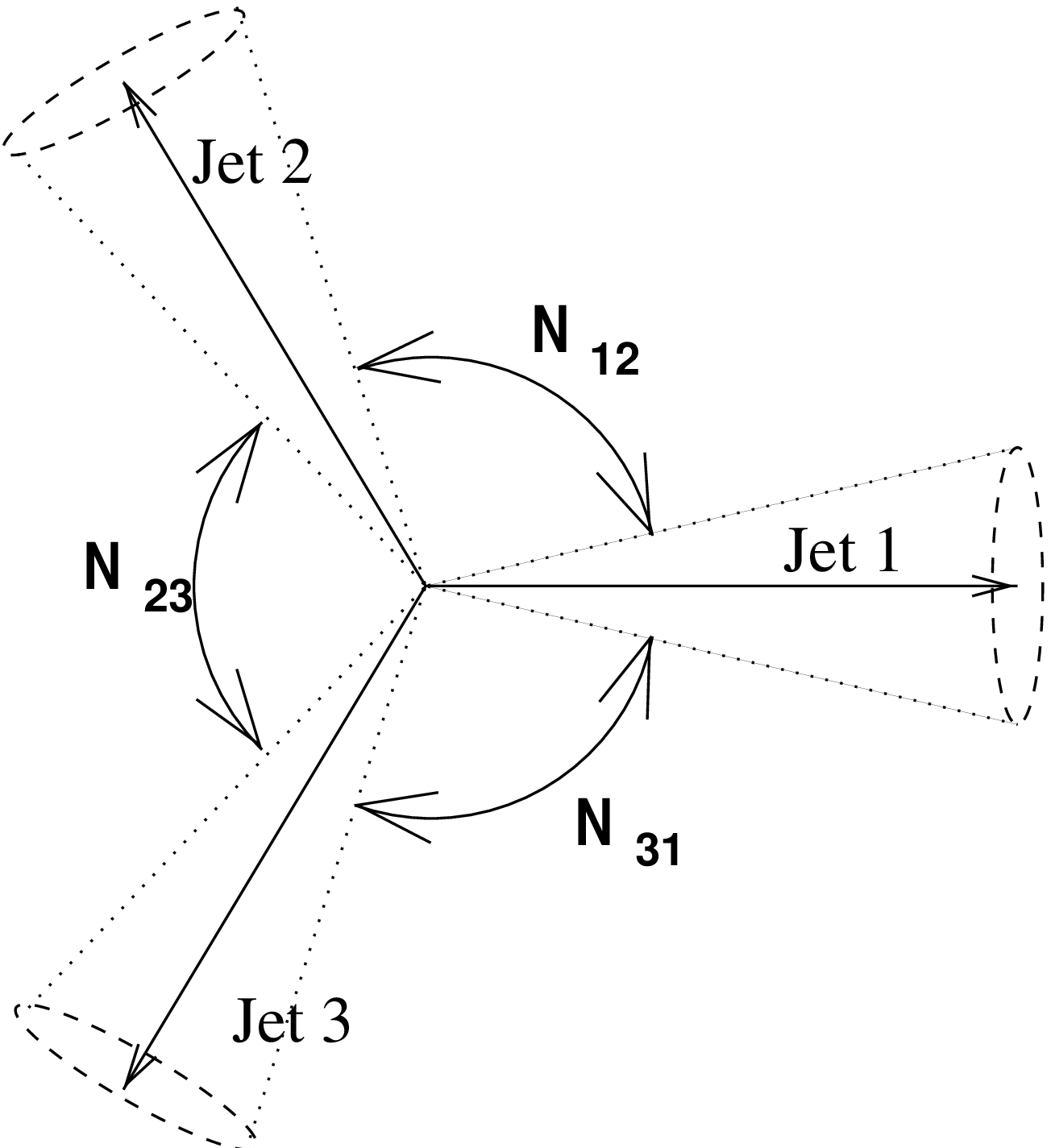}
\hspace{1cm}
\includegraphics[width=.35\textwidth]{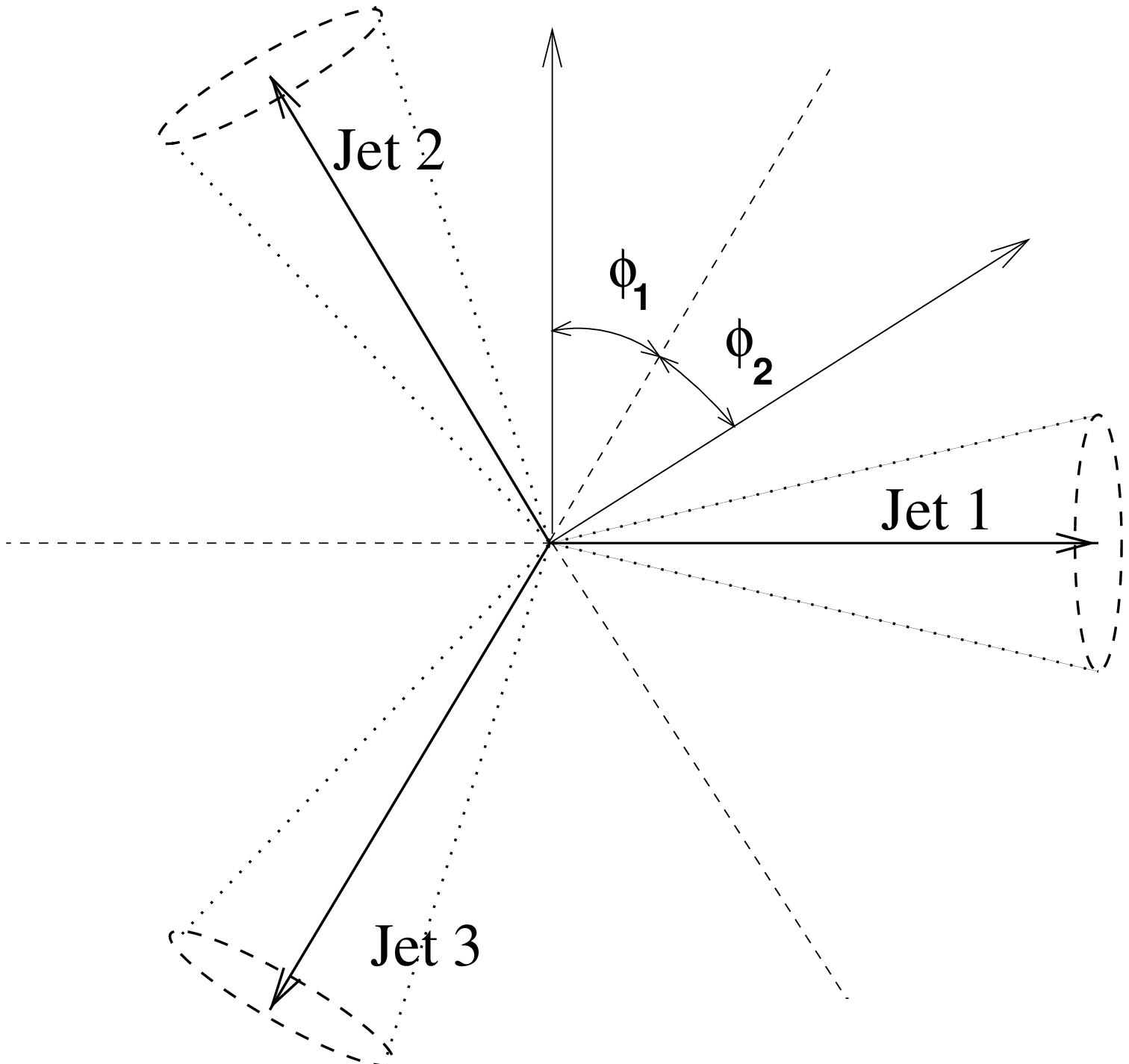}
\caption[]{Definitions of numbers of particles in inter-cone gaps (left) and angles
 relative to gap bisector (right).}
\label{fig:gaps}
\end{center}
\end{figure}

Three jet events are selected at jet resolution
parameter, 0.05, using the \textsc{Jade} algorithm~\cite{jade}, 
 with inter-jet angles within
$\pm 30^{\circ}$ from the symmetric Mercedes topology.
Particle momenta are projected on to the event plane defined
by the two most energetic jets, and then all the angles
(measured from most energetic jet) are rescaled so as to align
jets at 0$^\circ$-120$^\circ$-240$^\circ$ for
uniformity in event-to-event comparison.
To quantify the inter-jet gap angle, two definitions are used:
the minimum opening angle of the particles measured 
from the bisector in each gap, $B_{ij} = min(\phi_1^{ij},\phi_2^{ij})$, 
({\em{B-angle}}, figure~\ref{fig:gaps}), 
and maximum separation angle between adjacent particles in each gap, $S_{ij}$, 
({\em{S-angle}}). 
To minimise fragmentation bias,  
cones of half angle 15$^{\circ}$
around the jet-axis are removed in the definition of the gaps. 
Assigning jets 1, 2 to primary quarks, and
introducing the generic variable for the gap between jets $i$, $j$
as $\theta_{ij}$ $(i,j=1,3)$, where $\theta_{ij}$ may be the
 number of particles in the gap, $N_{ij}$, $B_{ij}$
(see figure~\ref{fig:gaps}) or $S_{ij}$, the  
{\em gap asymmetries} are defined as:
A$_{12}$ = $(-\thfs+\thst+\thtf)/(\thfs+\thst+\thtf)$,
and cyclically for the other gaps.
Reduced colour flow and thereby larger separation for \textsc{Cse} 
in gaps 23, 31 (together referred to as qg) with respect to gap 12, 
should thus make A$_{12}$ peak more strongly at positive values 
for \textsc{Cse} than for \textsc{Coe}. The angular asymmetries,
in particular that using the bisector angle, are found to be more sensitive
in separating \textsc{Cse} and  \textsc{Coe} contributions than those using
 $N_{ij}$.

\begin{figure}[hbtp]
\begin{center}
\includegraphics[width=.46\textwidth]{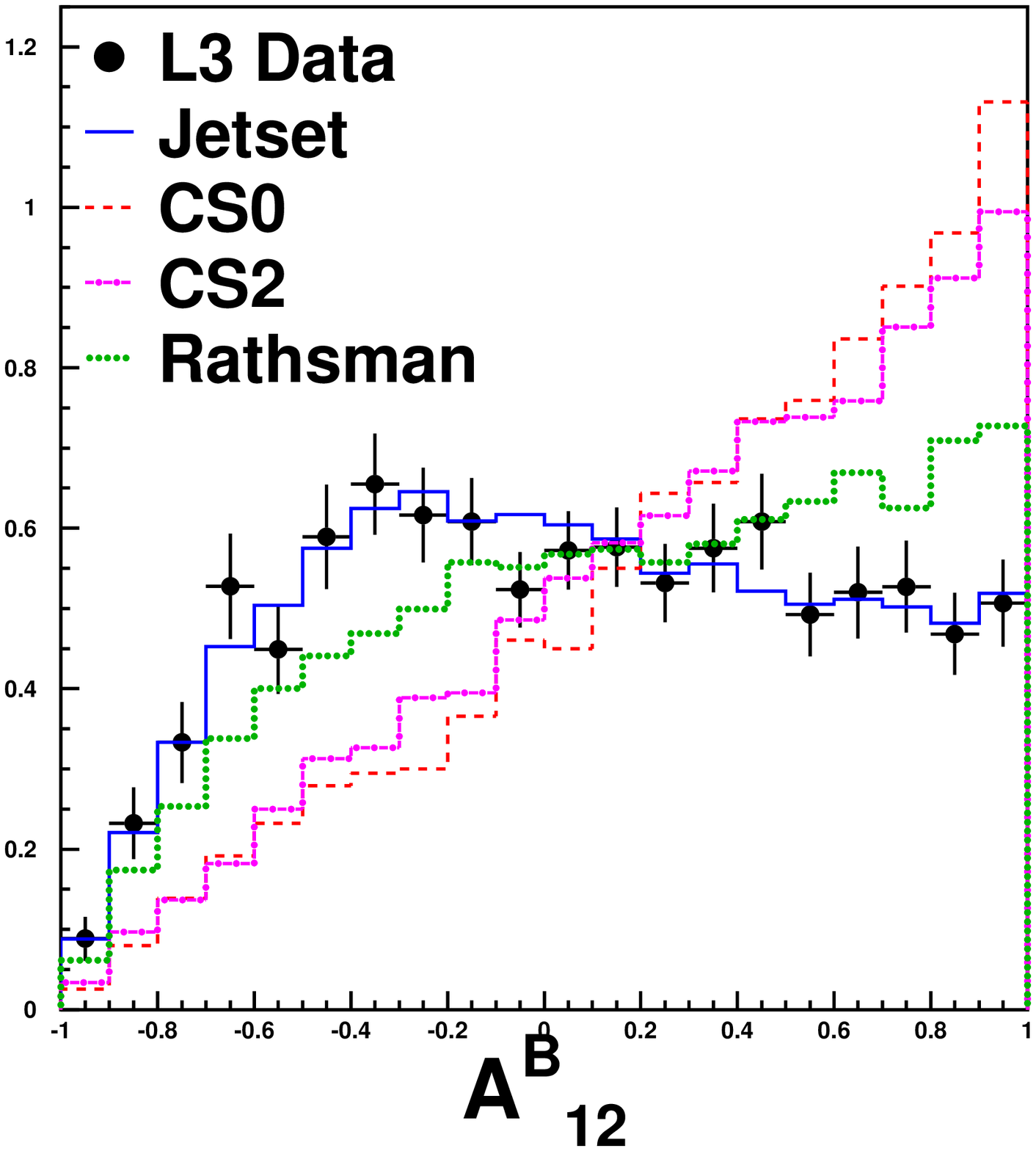}
\hfill
\includegraphics[width=.46\textwidth]{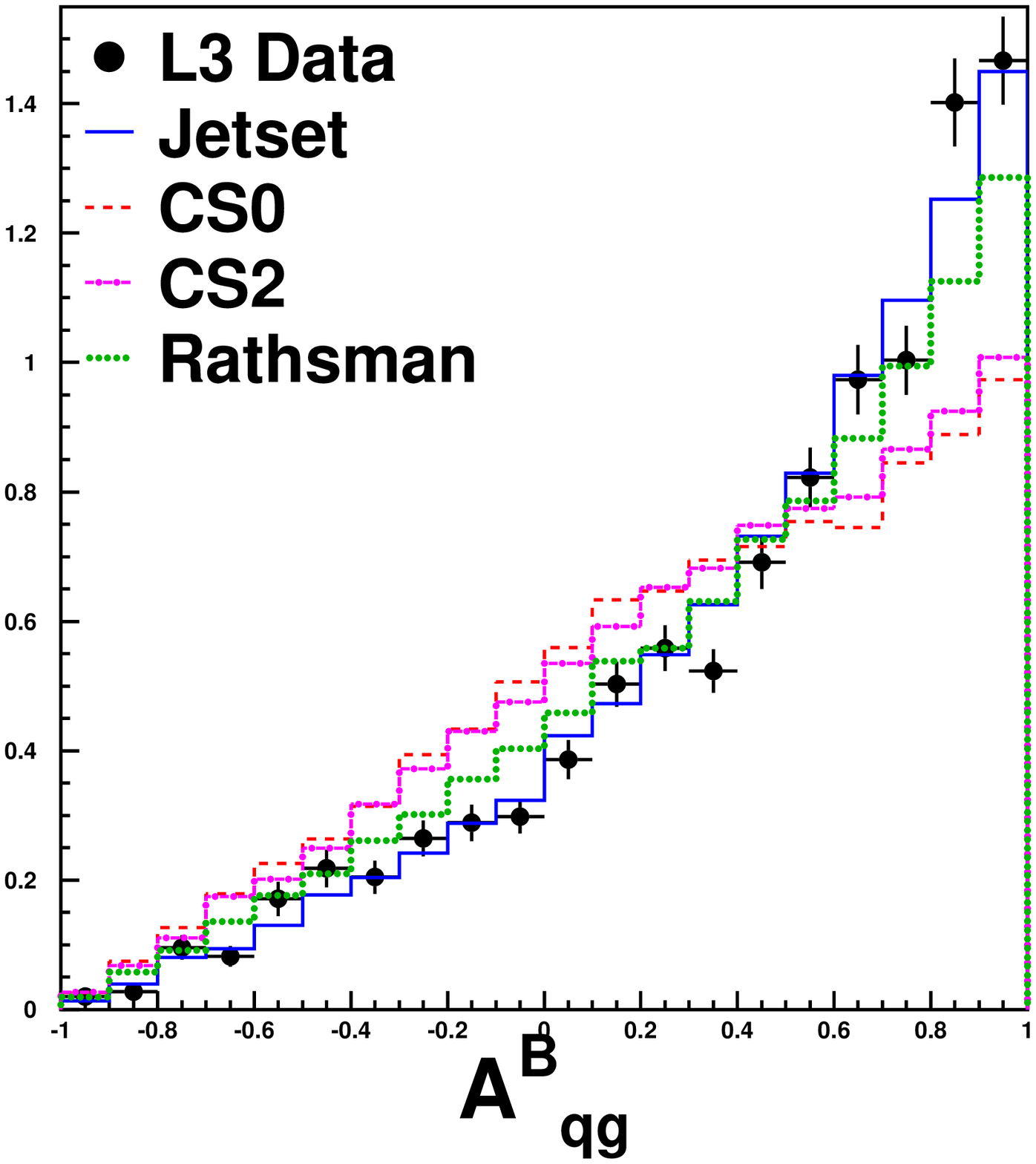}
\caption[]{Minimum bisector angle gap asymmetries for gaps 12 and qg.}
\label{fig:ab}
\end{center}
\end{figure}

\section{Results}
The analysis is performed with 2 million hadronic Z decay events
recorded by the L3 detector during 1994-95.
In order to distinguish quark jets from gluon or colour singlet jets, 
quark jets are tagged by demanding that the
b-tag discriminants~\footnote{The jet b-tag discriminant 
 is defined as: ${\rm B}_{jet} = -\log_{10}{\rm P}$ where ${\rm P}  ={\rm P}_i^n
 \sum_{j=0}^{n-1}(-\ln{\rm P}_i^n)^j/j!$ and ${\rm P}_i^n~=~\Pni$. Here, ${\rm P}_i$
 is the probability that
 the ith particle in the jet originates at the primary vertex.} 
 of jets 1 and 2 are above a certain cut-off ($\B_1 = 1.25 $) 
and that of jet 3 is below a second cut-off ($\B_2 = 1.5 $).   
This optimised selection
tags 2668 events with a gluon purity of 78\%. The asymmetries are calculated using 
 calorimetric clusters with at least 100 MeV in the electromagnetic calorimeter and either
 at least two crystals hit or at least 900 MeV in the hadron calorimeter, or, no
 energy deposit in the electromagnetic calorimeter, and more the 1800 MeV in the hadron
 calorimeter. As a cross-check, the analysis was repeated using only charged tracks with 
 $p_t > 100$ MeV.
The $B_{ij}$ asymmetry distributions, corrected for detector
effects as well as flavour composition,
are compared to different models,
all normalized to unit area, in figure~\ref{fig:ab}.
 The bin-by-bin correction
 factor typically lies in the range of $\pm 20 \%$. The fractional bin-by-bin systematic
  errors 
 were estimated by using the  \textsc{Durham} ($k_{\bot}$)
 algorithm~\cite{kt} with $y_{cut} =0.01,0.02$ instead
 of the JADE one, giving $\sim$ 4-7\%, and by changing the b-tag cuts so that the gluon purity
 varies by $\pm 10$\% resulting in an error of $\sim$ 3-8 \%.
 These errors are added in quadature. Changing the
 jet cone angle from 15$^{\circ}$ to 20$^{\circ}$ is found to have no effect on the
 angular asymmetries.  
The data are consistent with \textsc{Coe} as modelled by \textsc{Jetset}.
The probability for having a $\chi^2$ greater than the
observed value between data and  the \textsc{Rathsman} model
 with the default colour reconnection
parameter value $R_0 = 0.1$ is
${\cal{O}}(10^{-6})$. Fits to data with admixture of 
a fraction of \textsc{Cse}
to \textsc{Coe} are consistent with zero admixture.
Making a combined fit to the asymmeties between gaps 12 and qg, 
the upper limit of the fraction of \textsc{Cse} present in data,
obtained with the \textsc{CS0} or \textsc{CS2} models,
 is estimated
to be between 7\% and 9\% at 95\% confidence level. A similar fit for the
 parameter  $R_0$ of the Rathsman model, using the $B_{ij}$ asymmetries,
 yields a 95\% confidence level upper limit on the former of 0.0093.
 This implies a very small mass shift for the W due to 
 colour reconnection effects, in this model, of a few MeV.  For the
 default value, $R_0 = 0.1$, the mass shift for decay of W-pairs into 
 four jets is about 65 MeV~\cite{rath}. It is clearly then of interest to perform a
 similar analysis using other colour reconnection models (for example
 \textsc{Ariadne}~\cite{Ariadne}) applicable both to Z and W-pair decays. Since
 the fraction
 of colour singlet exchange expected, on the basis of the {\textsc{Tevatron}
 measurements, is only 5-10\%, after allowing for the effect of gap
 survival probability, the present analysis is not sufficiently 
 sensitive to confirm or exclude
  a similar effect in hadronic Z-decays. A LEP
 combination, and/or an analysis with higher statistics using asymmetric
 3-jet events is then desirable.

\section*{Acknowledgments}
I would like to thank Sunanda Banerjee, Swagato Banerjee and Dominique Duchesneau
 for their contributions to the work described here. Thanks are
 also due to Elisabetta Barberio for pointing out the interest of
 the angular asymmetries for colour reconnection studies.

\end{document}